# Optimal QoS-Aware Network Reconfiguration in Software Defined Cloud Data Centers

Mohammad Mahdi Tajiki
Electrical and Computer Engineering
Tarbiat Modares University
Tehran, Iran
Mahdi.Tajiki@modares.ac.ir

Behzad Akbari, Nader Mokari
Electrical and Computer Engineering
Tarbiat Modares University
Tehran, Iran
{b.akbari, mokari}@modares.ac.ir

*Abstract*— Software-defined networking (SDN) as a new paradigm for networking provides efficient resource reallocation platform in emerging cloud data center networks. The dynamic nature of cloud data center network's traffic, as well as the existence of big flows make it necessary to periodically reprogram the network through the SDN controller. Therefore, it is critical for network researchers to minimize the side-effects of network reconfiguration. In this way, the most challenging issue is the number of rerouted flows that affect the network stability and QoS parameters. As a result, dynamic reconfiguration of the network with minimum overhead (i.e. minimum flow rerouting) is an interesting problem in SDN-based resource reallocation. In this paper, we mathematically formulated the resource reallocation problem as an optimization problem with minimum network reconfiguration overhead subject to QoS requirements of the applications' flows. In order to reduce the time complexity of solving the optimization problem, a forwarding table compression technique is devised making the proposed scheme an efficient resource reallocation method which can be used as an application on top of the SDN controller. Our Experimental results show that the proposed scheme decreases the network reconfiguration overhead dramatically while meeting the QoS constraints. Since the reconfiguration overhead of the proposed scheme is low, the controller can reallocate the resources more frequently based on the network condition. We also studied the impact of the proposed network reconfiguration scheme on packet loss and delay in the network. The results show that the proposed approach outperform the conventional methods.

## 1 Introduction

The emergence of new bandwidth demanding applications such as video, cloud and over-the-top (OTT) services and also smart devices (e.g. Smartphones and Tablets) makes an evolution in computer network's traffic pattern. Cloud computing is developed to support the requirements of these new services and devices. As a consequence, data centers have evolved from a room packed with workstations to large-scale warehouses including hundreds of thousands of servers. In these new large-scale data centers, network infrastructure and protocols play a critical role. On the other hand, the quick acceleration of the Internet evolution has made the weaknesses of its current architecture and protocols so visible that they can be no longer masked by simple over-dimensioning network infrastructures [1]. These issues motivate the network researchers to a revolution from traditional networks to programmable networks. In this way, the state-of-the-art paradigm called software-defined networking (SDN) gives a new perspective to the network management in which the control plane is centralized and provides a global knowledge to the network managers. In brief, to address the mentioned issues SDN (which is a programmable network technology that is proper for dynamic and high-bandwidth network applications) is proposed.

Lots of today network applications such as media streaming, online gaming, cloud services and etc. require predictable, steady network resources with strict QoS requirements. In software-defined networking, OpenFlow provides flow level programmability that can be leveraged to program the network according to QoS requirements of the applications and also network traffic condition, dynamically. As a result, QoS-aware network reprogramming or reconfiguration plays an important role in traffic steering in multi-service SDN-based networks. In this type of resource reallocation, links are assigned to network traffic flows based on their QoS requirements and also traffic engineering objectives. In this way, there are several important issues that make flow level resource allocation more challenging: 1) big flows and resource partitioning in the network, 2) existence of burst and dynamic traffic, and 3) various traffic classes with different requirements. One of the most important side-effects of network traffic dynamicity and big flows is resource partitioning. Suppose there are two 10 Gb/s path from node A to node B. Each of them has 500 Mb/s free bandwidth and there is a flow with 800 Mb/s rate from node A to node B. Due to the resource partitioning, the flow cannot be routed properly, however, the network has 1 Gb/s free bandwidth capacity. In order to overcome the effects of resource partitioning, the flow routing must be done with a global view of the network and also considering its impact on all of the other flows. In other words, in order to route a big flow, we may need to reroute a number of other flows due to resource partitioning. In this way, the static network configuration methods are inefficient. Besides, with the existence of big flows and dynamic traffic in the network, it is necessary to reprogram the network frequently which may result in instability and undesirable impact on QoS of the flows. Therefore, these network characteristics make the problem of network reconfiguration with minimum side-effect and overhead (i.e. minimum changes in forwarding tables) more interesting in SDN networks. It should be mentioned that the network reconfiguration overhead (or precisely the flow rerouting overhead) is dependent on the number of rerouted flows. Increasing the number of rerouted flows not only may result in the network instability but also it may increase the packet loss and end-to-end delay.

In brief, the most important differences of our work with the traditional approaches are as follows: 1) lots of traditional approaches focus on the routing of new flows while QNR (the proposed QoS-aware Network Reconfiguration) focuses on the rerouting of existing flows. 2) Since the proposed algorithm considers the effect of flows on each other, it can handle resource partitioning. 3) Despite traditional algorithm, QNR can be used along with any other routing algorithm. 4) QNR focuses on network reconfiguration overhead and reroutes the flows in a way that minimizes the network reconfiguration overhead.

In this paper, the QoS-aware network reconfiguration problem with minimal changes in the forwarding tables is mathematically formulated. In order to overcome the side-effects of resource partitioning in the network, the proposed scheme is considered as a multi-path approach in which two flows from an identical source and destination can be routed through different paths. Additionally, it reallocates the resources in a way that imposes the minimum overhead to the network in reconfiguration phase. Since there are various traffic classes in the network, the proposed approach supports $k$ different traffic classes with various QoS requirements. Briefly, a dynamic and efficient network reconfiguration scheme is proposed in which flow bandwidth requirements are guaranteed and the network reconfiguration overhead is minimized. Besides, in order to reduce the time complexity of network reconfiguration, we have also proposed an approach to achieve a near-optimal solution. In this way, a forwarding table compression technique is exploited to make the scheme more efficient. Moreover, we made a tradeoff between time complexity and the accuracy of the solution by setting a variable compression rate for the flows. Therefore, the exchanged information between a special application with another application in another server can be considered as a flow while all traffic from one data center to another one can be considered a flow, too.

The main contributions of this paper are summarized as follows:
- Mathematically formulation of the QoS-aware network reconfiguration problem with minimal changes in the forwarding tables.
-  Addressing the resource partitioning problem in flow routing.
- Using a forwarding table compression technique to reduce the complexity of network reconfiguration.

The rest of this paper is organized as follows: related QoS-aware resource allocation algorithms are described in Section 2. The assumptions and basic information that are required to define our scheme are discussed in Section 3. Section 4 and Section 5 describes the main problem of this work as well as our solution. Section 6 compares the proposed solutions with a traditional approach. Finally, the paper is concluded in Section 7.

2 RELATED WORK

There are numbers of research works on optimal routing traffic in OpenFlow-based networks. Shetty *et al*. [2] present a network-aware resource reallocation technique, in which they use the network topology characteristics of the data center to minimize the maximum latency in communication between VMs (Virtual Machine). They incorporate the resource heterogeneities by including the computational and communication requirements in the proposed technique. The work [3] proposes a unified approach that integrates virtual machine and network bandwidth provisioning. They solve a stochastic integer programming problem to obtain an optimal provisioning of both virtual machines and network bandwidth when demand is uncertain. The mentioned paper is an inter-data center protocol (it is composed of cloud service providers and ISPs) and is not proper for intra-data center traffics. Some other works such as [4], [5], and [6], minimize the delay and packet loss by rerouting video flows. Hence, the end-to-end delay of each flow is guaranteed to be less than a predefined threshold. Hilmi *et al.* [7] present an analytical framework for optimization of forwarding decisions at the control layer to enable dynamic QoS over OpenFlow networks. They discuss the application of this framework to QoS-enabled streaming of videos with two QoS levels. They propose a dynamic QoS routing as a constrained shortest path problem and solve it by considering a two layer QoS problem. In this way, the base layer is treated as a level-1 QoS flow, while the enhancement layers are treated as best-effort flows. The authors of [8] propose a two-phase flow embedding approach with an iterative traffic engineering algorithm to address the resource reallocation problem in multimedia IP communication systems. Some other works such as [9] focus on providing QoS for VoIP (Voice over IP) traffic and simultaneously optimizing the power efficiency.

Civanlar *et al.* [10] propose an architecture to support QoS flows in an OpenFlow environment. They primarily focus on the analytical framework for optimization of the QoS flow routing. They also describe the control layer messaging between the controller and forwarder to set up queues, detect congestion and reroute traffic streams that require QoS. The work [10] proposes an algorithm for video streaming that minimizes the packet loss along with the path hops. The mentioned resource reallocation approaches can support three different traffic classes.

The main purpose of most of the routing algorithms is to aid the fast selection of a proper route, which should be adaptive, flexible, and intelligent for efficient network management. But usually, only one constraint or single mixed constraint is considered. The focus of [11] is to develop a genetic algorithm (GA) based routing algorithm that satisfies multiple constraints of the multimedia applications. Hence, they propose a heuristic called Multi-constraint QoS Unicast Routing Using Genetic Algorithm (MURUGA), which incorporates multiple constraints required by multimedia applications to find the feasible path satisfying the constraint requirement. They exploit the GA to optimize the delay and packet loss. Similarly, [12] defines the *path weight* as a QoS measure and minimize it with the aid of an ant colony system (ACS). It should be mentioned that due to the nature of ACS and GA, the authors of [11] [12] propose a function which is minimized by heuristic methods instead of defining a mathematical formulation.

Santosh *et al.* [13] define *critical links* as links specifying the maximum throughput of each path. They focus on these links and select a path based on its *critical link weight*. They propose a constraint-based routing algorithm for MPLS networks. The algo-rithm uses both bandwidth and delay constraints. It means that the delay of the selected path by the algorithm is less than or equal to the delay constraint value. Therefore, the residual bandwidth of the links must be equal to or greater than the bandwidth constraint value. Besides, in [13] the best path is computed based on avoiding critical links to reduce call blocking rate. In addition, the paths which are not satisfying the bandwidth and delay constraints are deleted to reduce the complexity of the algorithm. Thereafter, the algorithm uses the Shortest-Path algorithm to reduce path length. The reference [14] follows a similar approach by minimizing $\frac{traffic\ load}{capacity}$ of each link.

The Ghosh scheme [15] prevents flows from violating a predefined threshold by minimizing end-to-end delay and maximizing network throughput. They investigate a semi-centralized design to achieve scalability by distributing the computation across multiple tiers of an optimization machinery. Ongaro [16] exploits the use of SDN in conjunction with OpenFlow to manage differentiating network services with QoS. Initially, they define a QoS Management and Orchestration architecture that allows them to manage the network in a modular way. Then, a seamless integration between the architecture and the standard SDN paradigm is defined. Precisely, they use Integer Linear Programming (ILP) to formulate the network traffic routing in a way that minimizes delay and packet loss.

The aforementioned works focus on intra-AS (Autonomous System: a network under a single administrative entity) resource reallocation [23-25] but inter AS resource reallocation is also important. SDN has been recently applied in inter-DC traffic management. In this way, the computational cost of the centralized traffic engineering algorithm is an important issue because of the fact that the efficient and fast response of the algorithm is crucial to the practicality of SDN-based approaches. In [17], the authors present a utility-optimization-based joint-bandwidth reallocation for inter-DC communication with multiple traffic classes that handles priorities between traffic classes in a soft manner and explicitly considers the delay requirement of Interactive flows. They apply approximation techniques to solve the LP problem and obtain fast and accurate approximations. Finally, it should be mentioned that none of the reviewed algorithms focused on the network side-effects (overhead) of resource reallocation.

3 NETWORK MODEL AND ASSUMPTIONS

In this paper, we assume an SDN-based data center network using OpenFlow protocol to dynamically program the network. With the purpose of separating the forwarding element from the network intelligence, each switch forwards the flows based on a forwarding table. The controller allocates the resources to each flow by the scheduling of forwarding tables. In other words, the controller reconfigures the network by updating forwarding tables of the switches. The controller can obtain the current flow matrix and network topology via querying the switches. At this point, the network is reconfigured in a way that minimizes the flow table changes and simultaneously meets QoS constraint, e.g., the bandwidth of network applications is guaranteed.

Consider a network with $n$ OpenFlow-enabled switches, we represent the network topology with a matrix $B_{n \times n}$ where $B_{ij}$ determines the bandwidth of the link from the switch $i$ to the switch $j$. The number of flows in the network is $p$. The *new routing matrix* ($A_{n \times n \times p}$) specifies the path selected for each flow, e.g., if $A_{ij}^f \in \{0,1\}$ is equal to 1 then the flow $f$ crosses the link $i \rightarrow j$. The matrix $A0_{n \times n \times p}$ is the *current routing matrix* which is going to be reprogrammed. Suppose there are multiple classes of traffic in the network. The *flow requirement matrix* $C_{1 \times p}$ specifies flows requirements based on the corresponding class. The $i^{th}$ row of *flow requirement matrix* that defines the bandwidth requirement for the flows. It could be computed based on current flow requirements, a predefined threshold, or prediction methods. Table I summarizes the notation used in this paper.

TABLE I. NOTATION DEFINITION.

| Definition | Symbol |
| --- | --- |
| Number of switches | $n$ |
| Number of flows | $p$ |
| Matrix of link bandwidth | $B_{n \times n}$ |
| Flow requirement matrix | $C_{1 \times p}$ |
| Matrix of source switch of flows | $s_{1 \times p}$ |
| Matrix of destination switch of flows | $d_{1 \times p}$ |
| Current routing matrix | $A0_{n \times n \times p}$ |
| Link utilizaztion ratio | $\mu$ |
| **Problem Variables** | |
| New routing matrix | $A_{n \times n \times p}$ |

It should be mentioned that along with QNR there is a primary routing algorithm (common routing algorithm like ECMP) which routes the new flows in real time. In contrast to the existing and common routing algorithms, QNR reallocates the resources in a completely separated manner (e.g., it reallocates the resource if the network gets congest or if a predefined time interval is met). Figure 1 provides a comprehensive view of the network in our design. Our proposed algorithm has two following modules:

1) <u>Primary routing</u>: A common routing algorithm which runs when there is a new arrival flow.
2) <u>Secondary routing</u>: the proposed QNR algorithm which runs when there is a network congestion or the predefined time interval is elapsed.

In contrast to the primary routing module, determining the path of new arrival flows, the secondary routing module reroutes all flows to avoid the congestion and improve the network throughput.

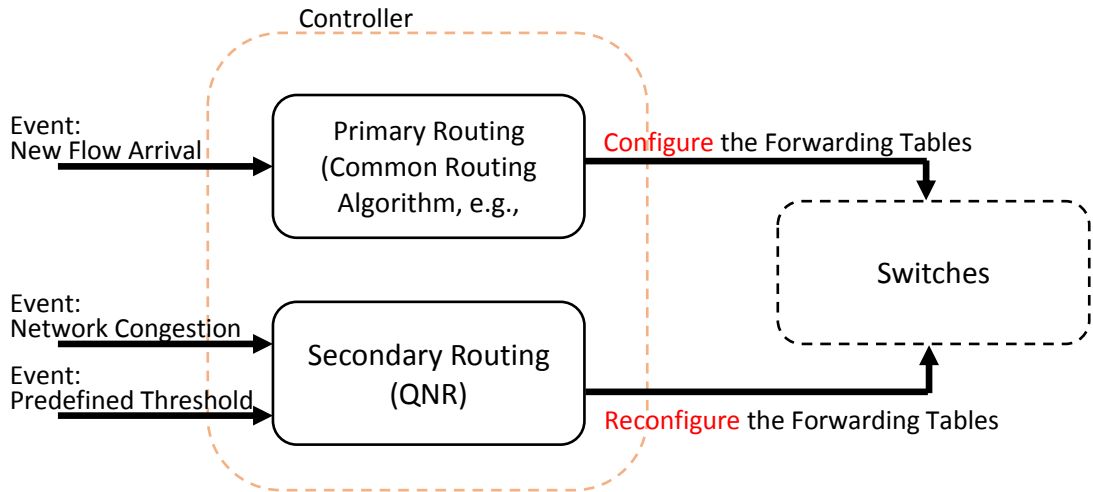

FIGURE 1. PROPOSED ARCHITECTURE.

As can be seen in Figure 1, if a new flow arrives at a switch, the controller uses common routing algorithm (such as ECMP or the algorithm introduced in [3]) to route the flow. Thereafter, the controller configures the switches that are in the selected path by sending them controlling messages. On the other hand, if the network gets congested (or if a predefined time intervals elapsed), the controller fetches the switches states and reconfigures them to reduce the congestion, improve the throughput and providing the requirements of the flows in the network. In brief, along with QNR, there is a primary routing algorithm which routes the new flows in real time. Therefore, we proposed an architecture that is completely different from the existing and common routing algorithms. In contrast to the existing algorithms, QNR reallocates the resources in a completely separated manner (e.g., it reallocates the resource if the network gets congest or if a predefined time interval is met).

4 PROBLEM FORMULATION

The main goal of this paper is to efficiently and dynamically reallocate resources in a way that 1) guaranties QoS requirements of different applications, 2) proactively prevents the resource waste and congestion, and 3) minimizes the network overhead in the reconfiguration process. The routing matrix should be calculated in a way that satisfies the mentioned constraints. To this end, *new routing matrix* ($A_{n \times n \times p}$) can be obtained such that minimizes the network reconfiguration overhead subject to the QoS constraint (2) and the flow conservation constraints (3-8). The problem formulation is as follows:

$$\underset{A}{\text{minimize}} \left( \sum_{i=1}^{n} \sum_{j=1}^{n} \sum_{f=1}^{p} \left| A_{ij}^f - A0_{ij}^f \right| \right) \quad (1)$$

*Subject to:*

$$\sum_{f=1}^{p} A_{ij}^f C_f \leq \mu B_{ij}, \quad \forall i,j \in \{1, \ldots, n\}, \quad (2)$$

$$\sum_{i=1}^{n} A_{is_f}^f = 0, \quad \forall f \in \{1, \ldots, p\}, \quad (3)$$

$$\sum_{i=1}^{n} A_{d_f i}^f = 0, \quad \forall f \in \{1, \ldots, p\}, \quad (4)$$

$$\sum_{i=1}^{n} A_{s_f i}^f = 1, \quad \forall f \in \{1, \ldots, p\}, \quad (5)$$

$$\sum_{i=1}^{n} A_{id_f}^f = 1, \quad \forall f \in \{1, \ldots, p\}, \quad (6)$$

$$\sum_{j=1}^{n} A_{ij}^f = \sum_{j=1}^{n} A_{ji}^f, \quad \forall f \in \{1, \ldots, p\}, \quad \forall i \in \{1, \ldots, n\} - \{s_f, d_f\}, \quad (7)$$

$$\sum_{j=1}^{n} A_{ij}^f \leq 1 \quad \forall i \in \{1, \ldots, n\}, \quad \forall f \in \{1, \ldots, p\}, \quad (8)$$

$$A_{ij}^f \in \{0,1\}, \quad \forall f \in \{1, \ldots, p\}, \quad \forall i,j \in \{1, \ldots, n\}, \quad (9)$$

Where (2) guaranties the link load to be smaller than a maximum predetermined threshold at which the left-hand side of the constraint calculates the sum of guaranteed bandwidth of all flows crossing a specified link. The right-hand side of it specifies the link allowed bandwidth. Fig. 2A is the visual illustration of (3) in which the flows are prevented from returning to the source switches. As it is mentioned earlier, $A_{is_f}^f$ is zero if and only if the flow $f$ crosses the link that connects switch $i$ to the source switch of $f$. For each flow, (3) forces the summation of $A_{is_f}^f$ (for all $i$) to be zero. In other words, none of the flows can cross the link between a switch to the source switch of that flow. On the other hand, (4) makes the flows to stay on the destination switches as depicted in Fig. 2B.

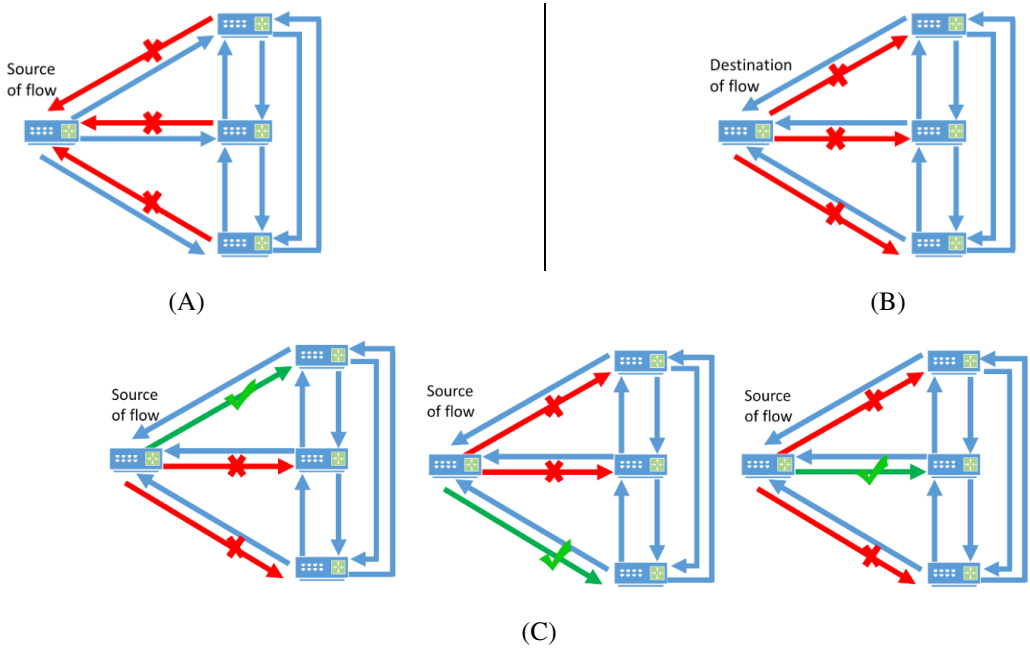

FIG. 2 VISUAL ILLUSTRATION OF CONSTRAINTS. (A) CONSTRAINT 3 (B) CONSTRAINT 4 (C) CONSTRAINT 5.

In order to force the flows to leave the origin switches and enter to the destination one, constraints (5) and (6) are considered, respectively. In this regard, (5) guaranties the flow to cross from exactly one of the source switch outgoing link (Fig. 2C) as well as (6) ensures the same manner for the incoming links of destination switch which is demonstrated in Fig. 3A. It should be mentioned that if a switch is neither source nor destination of a flow, the flow must leave that switch after it moves in. This restriction is applied by (7) via balancing the amount of traffic entered to the switch with the amount of traffic left it as depicted in Fig. 3B. Constraint (8) makes sure there is no loop in the *new routing matrix*. It prevents flows from returning to a switch that is met in the past. Finally, the objective function, (1), is to minimize the differences of *new routing matrix* and *current routing matrix*.

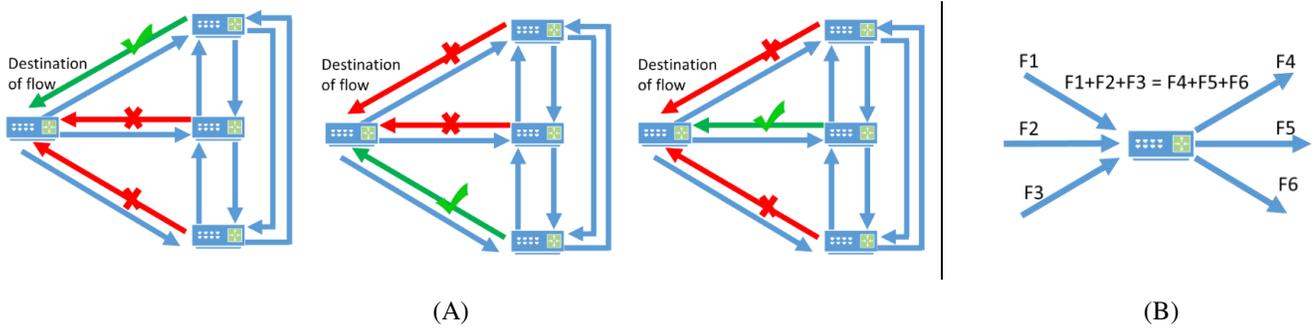

FIG. 3 VISUAL ILLUSTRATION OF CONSTRAINTS. (A) CONSTRAINT 6 (B) CONSTRAINT 7.

5 PROPOSED SCHEME

In order to solve the mentioned optimization problem, two approaches are proposed called QNR (QoS-aware Network Reconfiguration) and RQNR (Relaxed QNR). The first approach obtains the global solution when the problem is feasible. On the other hand, RQNR uses forwarding table compression techniques to decrease the number of active flows in order to reduce the computational complexity of network reconfiguration.

## 5.1 QNR

The absolute value of the objective function of the proposed optimization problem intensely increases the computational complexity. To overcome this difficulty, we remove the absolute value from the objective function and an equivalent function is introduced as follows

$$\sum_{i=1}^{n}\sum_{j=1}^{n}\sum_{f=1}^{p}\left(A_{ij}^{f} + A0_{ij}^{f} - 2 * A_{ij}^{f} * A0_{ij}^{f}\right). \tag{10}$$

To prove the equality, we can check all states similar to Table 2. Since the considered objective function, (1), is the sum of $n \times n \times p$ elements, each element contains four different states (0, 0), (0, 1), (1, 0), and (1, 1) where the first element is $A_{ij}^{f}$ and the second one is $A0_{ij}^{f}$, i.e. $(A_{ij}^{f}, A0_{ij}^{f})$. The results of both (1) and (10) is 0 for inputs (0, 0) and (1, 1). In (10), both inputs (0, 1) and (1, 0) result in value 1, similarly the outcome for (1) is 1, too. Therefore, the new objective function is a substitution of the original one. As it can be seen, the optimization problem is in form of Binary Linear Programming which is a branch of integer linear programming and it can be solved by available toolboxes like CVX, YALMIP, and CPLEX. Accordingly, we employ the CVX library (a package for solving ConVeX optimization problem) [18] to solve the mentioned problem.

*TABLE 2. EQUALITY OF EQUATIONS (1) AND (10).*

| $A_{ij}^{f}$ | $A0_{ij}^{f}$ | Output of (1) | Output of (10) |
|---|---|---|---|
| 0 | 0 | 0 | 0 |
| 0 | 1 | 1 | 1 |
| 1 | 0 | 1 | 1 |
| 1 | 1 | 0 | 0 |

There are some other optimizers which can solve the BLP efficiently if there is no negative coefficient in the objective function. In order to omit the negative coefficient in the objective function, we can add a new constraint. The new objective function and constraint must be as follows:

$$\sum_{i=1}^{n}\sum_{j=1}^{n}\sum_{f=1}^{p}\left(A_{ij}^{f} + A0_{ij}^{f}\right), \tag{11}$$

$$A_{ij}^{f} + A0_{ij}^{f} \leq 1 \qquad \forall\, i, j \in \{1, n\}\, f \in \{1, p\}. \tag{12}$$

It should be mentioned that we use the constraint (10) as our objective function.

## 5.2 RQNR

In this approach, in addition to the exploitation of the new objective function, (10), a forwarding table compression technique is proposed which decreases the computational complexity of the proposed scheme by reducing the number of active flows (Algorithm 1). For simplicity, all flows of a specified source and destination are called *SF flows*. There are lots of small flows in a network (flows with size less than 10KB) [19] which can be merged to reduce the time complexity of the solution. To this end, all *SF flows* that are smaller than a predefined lower band are merged. Note that the outcome (called a stream) must be smaller than an upper bound threshold, otherwise it breaks into two or more flows. Therefore, an increment in the lower band reduces the time complexity while it increases the optimality gap. Experimental results show that in a large number of flows, the compression reduces the number of active flows dramatically.

```
1.  for each flow f
2.      for all flows q after f
3.          if size(f)+size(q)< predefined-threshold
```

```
4.                    size(f)=size(f)+size(q);
5.                    Delete(q);
6.            End
7.        End
8.  End
```

*ALGORITHM 1. "FORWARDING TABLE COMPRESSION" PSEUDO CODE.*

## 6 RESULT ANALYSIS

In this section, the set-up of our simulation is explained and the performance and computational complexity of the proposed algorithm are discussed. In order to analyze the performance of QNR, the reconfiguration overhead, delay, and the packet loss are compared with the shortest-path algorithm. At end of this section, the computational complexity of the proposed algorithm is discussed.

### 6.1    Simulation Set up

Since fat-tree topology [20] is a scalable commodity data center network architecture that is universally adopted, it is employed to investigate the performance of the proposed models. It should be mentioned that this topology is very suitable and efficient for data center communication. Besides, the proposed analytical model is evaluated on the real network topology shown in Fig. 5. We use a university network traffic which can be found in [21]. Although the flows sizes are real, due to lack of information about the IP layout, we assign IP addresses to the switches randomly. All flows are considered to be UDP. Due to the high capacity of the fat-tree topology, the flows size are increased to properly show the impact of our scheme.

In reconfiguration phase, the controller updates the forwarding table of switches which is done by sending *OpenFlow messages* to each switch. In highly dynamic networks, the network reconfiguration takes place in short periods of time (e.g., each second) which makes these controlling messages challenging. To show the impact of it, the proposed scheme is compared with conventional approaches from the *network reconfiguration overhead* perspective. We consider both schemes (QNR and conventional) are QoS-aware.

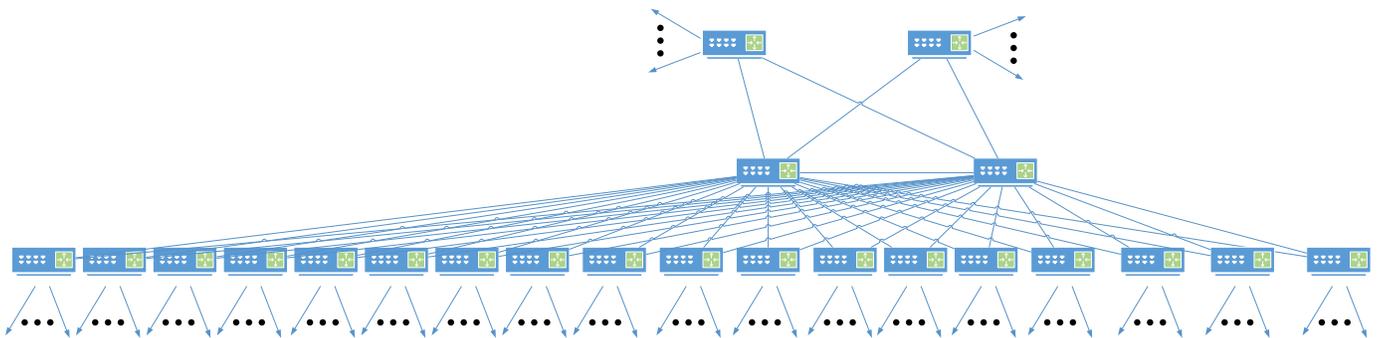

*FIG. 4  REAL NETWORK TOPOLOGY.*

### 6.2    Performance Analysis

In this subsection, the performance of the proposed algorithm is analyzed. To this end, QNR is compared with the shortest-path algorithm from reconfiguration overhead, delay, and packet loss.

#### 6.2.1    Reconfiguration Overhead

For the sake of simplicity, the number of changes in the elements of switches forwarding table is called SFTC (Switches Forwarding Table Change). In Fig. 5, SFTC versus the number of flows is depicted and in Fig. 6 the percentage of *the links with change (in the loading flows)* versus *the number of links cross the number of flows* is depicted (such consideration is to show the importance of the number of links in each topology). The topology used in Fig. 5A and Fig. 6A is a real data center network topology which is depicted in Fig. 4. Fat-tree topology with K=4 is used in Fig. 5B and Fig. 6B. Similarly, Fig. 5C and Fig. 6C are based on the fat-tree K=6 topology. The difference in these figures is about the traffic pattern which is explained by the parameter PL. It should be mentioned that if a flow changes its path from link $a$ to link $b$, then both $a$ and $b$ are considered as changed links (in Fig. 5 and Fig. 6).

As it can be seen in Fig. 5, increasing the number of flows raises the SFTC of traditional approaches dramatically while it is stable for QNR. It is because in the traditional approaches, increasing the number of flows increases the number of congested flows. To overcome this difficulty, traditional approaches have to reroute lots of flows. In contrast, in QNR, some special flows are only rerouted making the growth of SFTC small. The superiority of QNR over the traditional approaches is more remarkable when there exists big flows and resource partitioning within a highly dynamic network. Comparing Fig. 5B (fat-tree K=4) and Fig. 5C (fat-tree K=6), one can say that increasing the complexity of fat-tree topology reduces the SFTC in equal number of flows. This happens because by increasing the parameter K, the capacity of the network increases, consequently, the congestion rate and SFTC will be reduced. Based on our simulation, we understand that changing the traffic pattern (parameter PL) has not a direct effect on the SFTC.

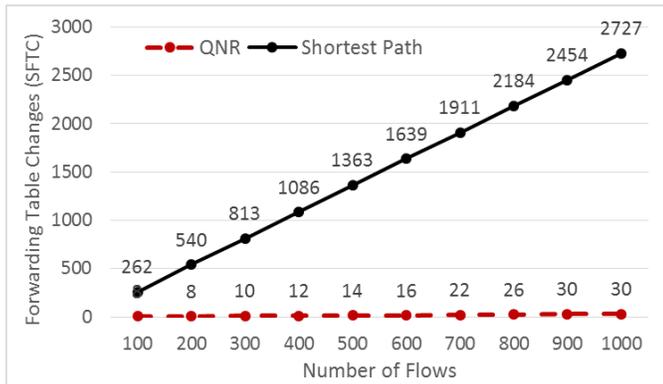

Fig. 5A) Real Network Topology

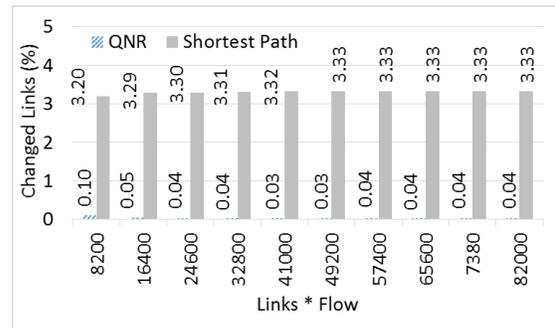

Fig. 6A) Real Network Topology

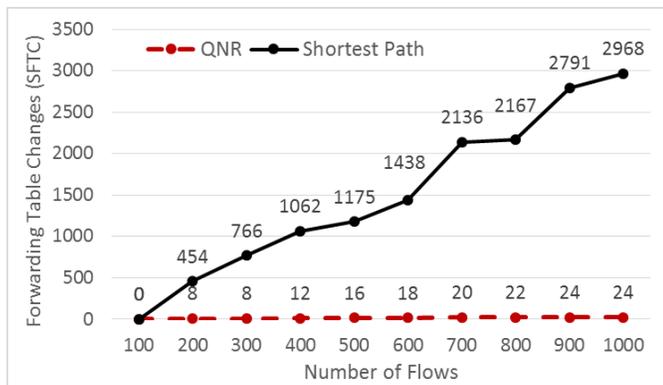

Fig. 5B) Fat-Tree K=4 PL=0.25

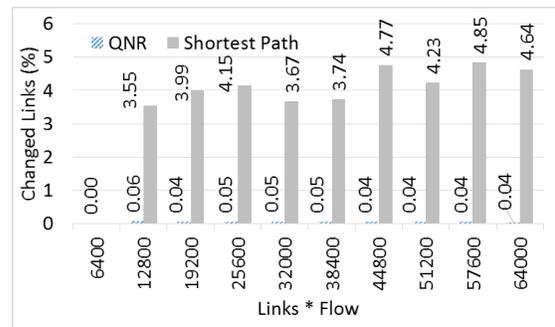

Fig. 6B) Fat-Tree K=4 PL=0.25

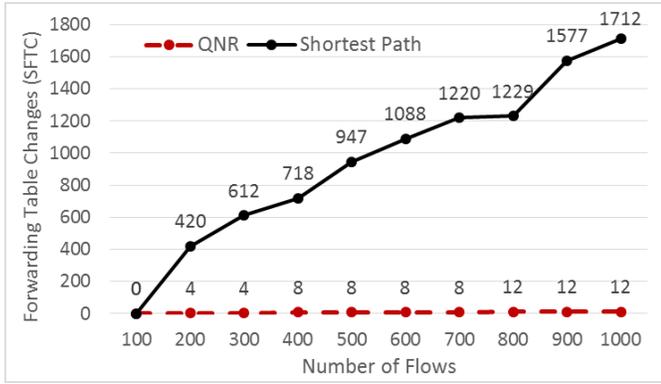
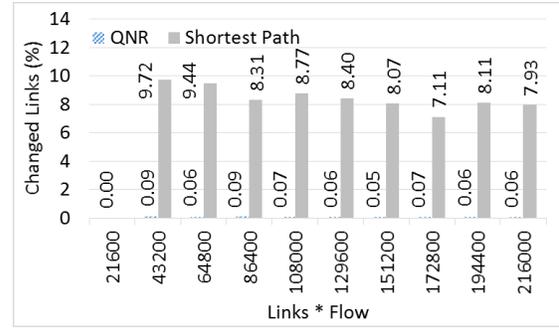

Fig. 5C) Fat-Tree K=6 PL=0.75

Fig. 6C) Fat-Tree K=6 PL=0.75

FIG. 5 OVERHEAD COMPARISON OF QNR AND CONVENTIONAL APPROACHES (SFTC).

FIG. 6 OVERHEAD COMPARISON OF QNR AND CONVENTIONAL APPROACHES (CHANGED LINKS).

One of the most important topics in cloud data centers is VM (virtual machine) migration. VM migration impose burst and big flows on the network. Therefore, the impact of big flows on our scheme is discussed in this part. In order to analyze the impact of big flows on the network, we exploit fat-tree topology (K=4, PL=0.25, and Number-of-Flows=400) with different number of big flows in the network. In this regard, the number of big flows are changed in the network and SFTC parameter is depicted. Figure 7 and Figure 8 show the number of forwarding table changes (SFTC) versus the number of big flows in the network. Increasing the number of big flows in the network increases SFTC in QNR, since by increasing the number of big flows, the flexibility of scheduler reduces. On the other hand, since shortest path is un-aware of number of rerouted flows in the network, increasing the number of big flows may increase or decrease SFTC (in most of cases it will increase SFTC).

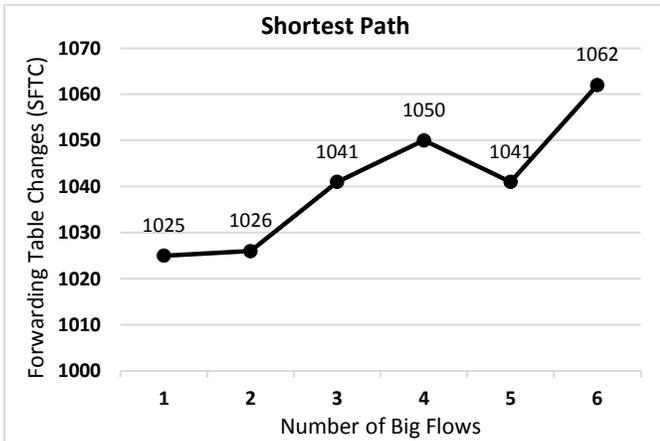
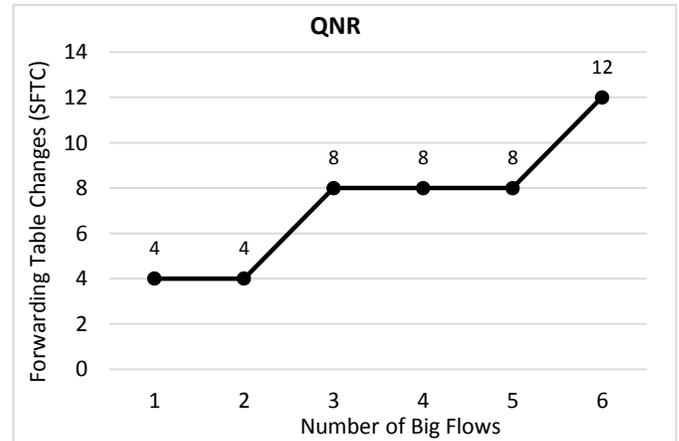

FIGURE 7. IMPACT OF BIG FLOWS (FAT-TREE K=4, PL=0.25)-SHORTEST PATH

FIGURE 8. IMPACT OF BIG FLOWS (FAT-TREE K=4, PL=0.25)-QNR.

### 6.2.2 Delay

There are two types of delay which are related to the network reconfiguration: a) end-to-end delay of control messages b) delay of updating the forwarding tables (FTs) in which the amount of these delays depend on the considered rerouting schemes. In this sub-section, we investigate the two mentioned delays.

- End-to-end delay of control messages

In this section, the end-to-end delay of control messages is mathematically formulated. This type of delay is a function of the number of rerouted flows. Suppose there are $n'$ flows that are rerouted via a new path. The propagation delay and the transmis-sion rate of all links are $t_{propagation}$ and $t_{transmission}$, respectively. The processing time of each switch (time of updating an element of FT) is $t_{proccess}$. The end-to-end delay of control messages is computed as follows:

$$E2E\ delay\ of\ control\ messages = t_{propagation} + t_{transmission} + \max\left((n'-1)*t_{transmission}, n'*t_{processing}\right) \quad (11)$$

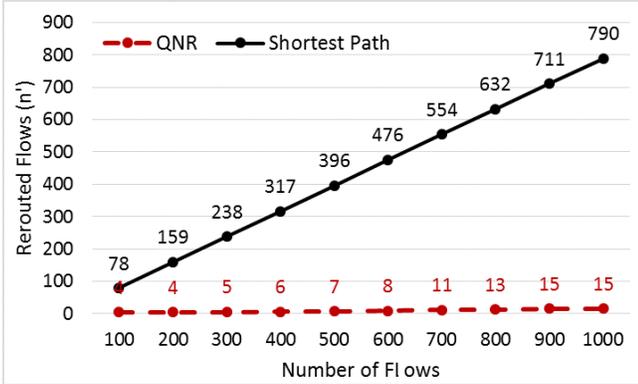

Fig. 9A) REAL NETWORK TOPOLOGY

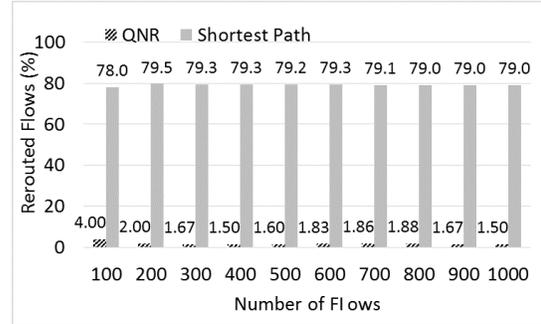

Fig. 10A) REAL NETWORK TOPOLOGY

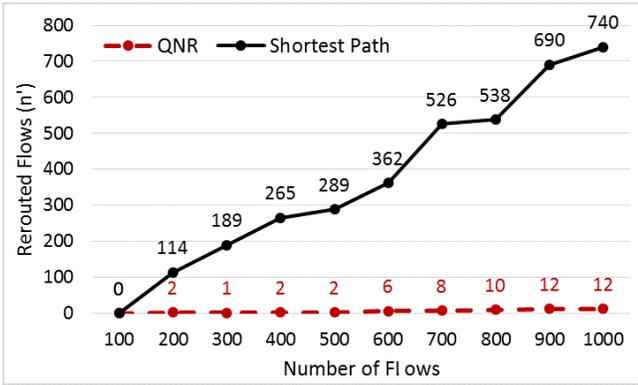

Fig. 9B) Fat-Tree K=4 PL=0.25

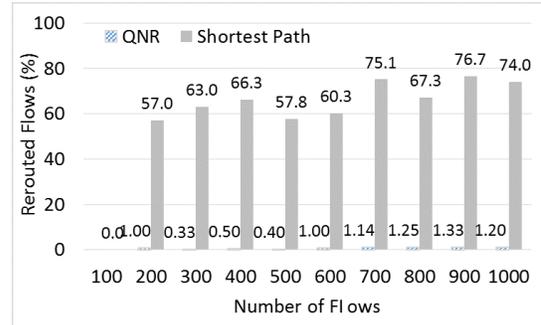

Fig. 10B) Fat-Tree K=4 PL=0.25

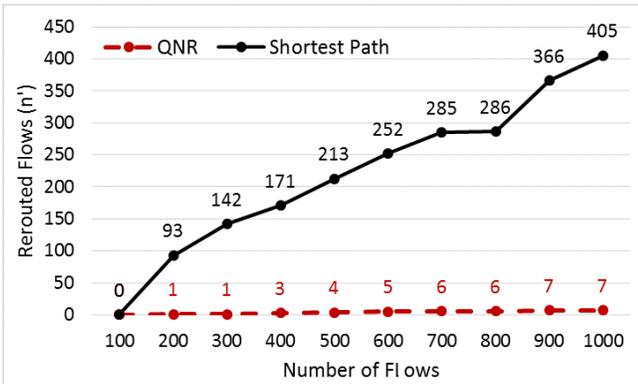

Fig. 9C) Fat-Tree K=6 PL=0.75

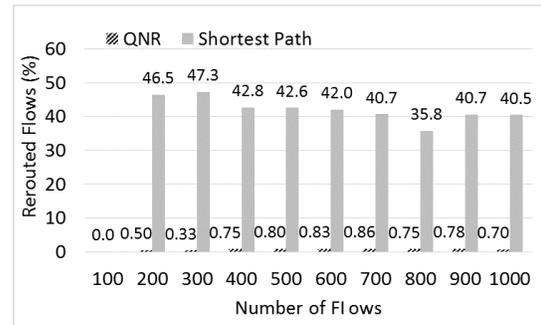

Fig. 10C) Fat-Tree K=6 PL=0.75

FIG. 9 NUMBER OF FLOWS THAT ARE REROUTED.

FIG. 10 PERCENTAGE OF FLOWS THAT ARE REROUTED.

As it can be seen from Eq. (11), by increasing the number of flows that are rerouted ($n'$) the end-to-end delay of control messages increases. In order to compare the Shortest-Path and QNR from this perspective, Fig. 9 and Fig. 10 are demonstrated. Fig. 9 shows the number of rerouted flows versus the total number of flows. Similarly, in Fig. 10 the percentage of the rerouted flows versus the total number of flows is depicted. The topology used in Fig. 9A and Fig. 10A is depicted in Fig. 4. Fat-tree topology with K=4 is used in Fig. 9B and Fig. 10B. In Fig. 9 it is shown that the number of flows with a new route is dramatically decreased in QNR compared with the Shortest-Path. The difference in these figures are about the topology and the traffic pattern which is explained by the parameter PL. Based on our results, increasing the parameter PL (the probability of leaving the rack) decreases the number of rerouted flows. Additionally, to clearly illustrate the impact of considering the *number of rerouted* flows as a performance parameter, Fig. 10 is depicted. In conventional approaches up to 79% of flows are rerouted while QNR reconfigures the network based on QoS requirements of flows in a way that at most 5% of flows are rerouted. Comparing Fig. 9B and Fig. 9C, we understand that increasing the parameter K in fat-tree topology, decreases the percentage of rerouted flows. This happens because increasing K raises the capacity of the network and consequently decreases the congested flows.

In all of our test cases, increasing the number of flows result in an increment in the number of rerouted flows. However, in Fig. 9B in the second step by increasing the number of flows the number of rerouted flows is decreased. It is because, some flows cross lots of switches and new route for them result in lots of FT entries change. Consequently, the algorithm reroutes two other flows instead of that flow. Therefore, although the number of rerouted flows is decreased in that step, the SFTC (network overhead) is increased.

- Delay of updating the forwarding tables

The delay of updating the *forwarding table (FT)* has two main parameters: 1) time of setting an FT element 2) number of elements in a FT that are going to be changed (NFTC: Number of FT Changes). The first one relies on the switch hardware, therefore it is independent of rerouting algorithm. However, NFTC is dependent on both the rerouting algorithm and the switch hardware. Fig. 11 illustrates the reconfiguration delay versus the total number of flows for QNR and Shortest-Path. Since NFTC value of switches are different in resource reallocation, the impact of QNR on the switch that has the maximum NFTC value is illustrated in Fig. 11. We consider it takes 1 ms for switches to change an element in their FTs and multiple requests in a FT take place sequentially. The topology used in Fig. 11A is depicted in Fig. 4. Fat-tree topology with K=4 is used in Fig. 11B. Similarly, Fig. 5C are based on the fat-tree topology K=6. The difference in these figures are about the traffic pattern which is explained by the parameter PL.

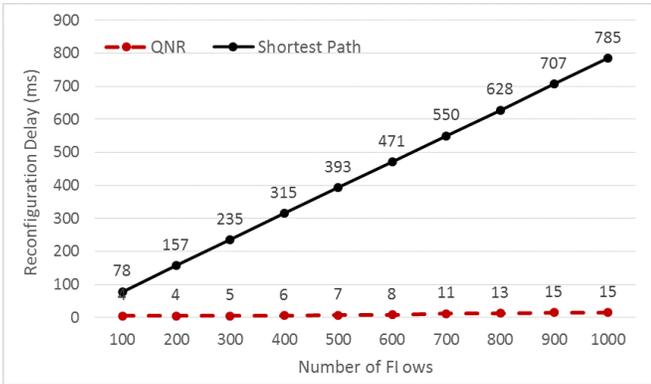

(A) REAL NETWORK TOPOLOGY

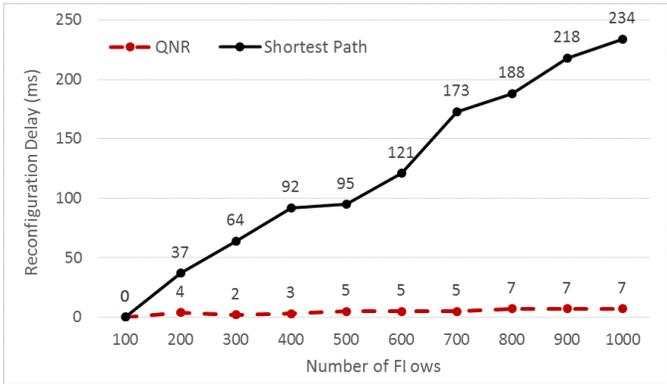

(B) Fat-Tree K=4 PL=0.25

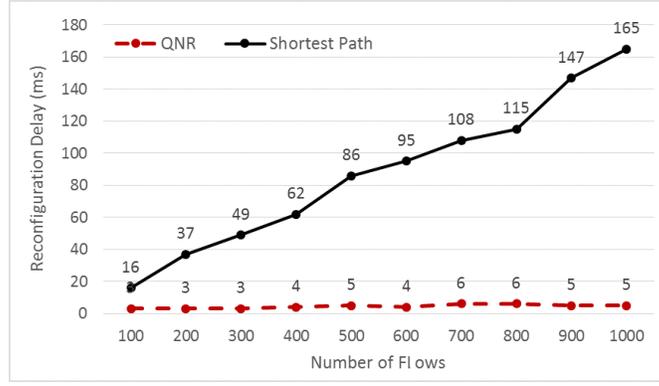

(C) Fat-Tree K=6 PL=0. 5

FIG. 11 MAXIMUM NUMBER OF CHANGES IN A SWITCH FORWARDING TABLE.

Based on the experimental result, the maximum NFTC value of QNR is dramatically lower than the Shortest-Path. Therefore, the delay of updating the FT is lower in QNR compared with conventional approaches like Shortest-Path. It should be mentioned that the length of paths are different and therefore in some cases rerouting of two flows may impose more entry changes when compared with reroute of three flows. In other words, although in some cases the number of rerouted flows is decreased, the NFTC is increased as it can be seen in Fig. 11C. The main reason that the reconfiguration delay of QNR is dramatically lower than the Shortest-Path is that the Shortest-Path imposes lots of entry changes in some few switches' FTs. Based on our results, the probability of leaving the rack (PL parameter) has a high impact on the reconfiguration delay. In contrast, the complexity of fat-tree topology (K parameter) does not has a direct impact on the reconfiguration delay.

### 6.2.3 Packet Loss

The impact of SFTC on the packet loss depends on two parameters: 1) selected path by routing algorithm 2) TCP retransmission timer. In the following these two parameters are discussed.

- The Impact of Selected path on the Packet Loss

A simple fat-tree topology (illustrated in Fig. 12) is selected to give a comprehensive description of the reason of packet loss in a SDN-based network reconfiguration. In this regard, we consider the bandwidth of all links is $x$ Gb/s. In order to simplify the problem, we ignore all flows that are in the network except two big flows (that are depicted in Fig. 13). The controller is semantically connected to all of the switches and it can communicate with all of them, however, this connection may be a direct connection (a physical link between the controller and the switch) or an indirect one (through the other switches). In Fig. 12, the switches that are directly connected to the controller are specified by bold dotted lines while the others are connected via narrow dotted lines. There are two big flows in the network that are currently routed via $S19 \rightarrow S11 \rightarrow S1 \rightarrow S5 \rightarrow S13$ and $S13 \rightarrow S5 \rightarrow S3 \rightarrow S9 \rightarrow S9 \rightarrow S17$. The guaranteed bandwidth for these flows are F1 Gb/s and F2 Gb/s, respectively. Suppose F1+F2 is greater than x, e.g., F1=0.9x Gb/s and F2=0.8x Gb/s (UDP Traffic).

To show the impact of routing algorithm on packet loss, we focus on rerouting of the red flow and the other flows (consisting the green one) are ignored. The propagation delay of links is specified by the matrix $D'_{1*20}$, however, by the purpose of simplifying the problem, all links delay are considered to be equal to $\tau$. As it can be seen in Fig. 12, the controller is directly connected to S1, S5, and S13 where their propagation delays are $\tau$. The propagation delay of controller messages to S19 is $3\tau$. As it was mentioned earlier, due to the resource partitioning, the network resources are reconfigured periodically; this period is denoted by T.

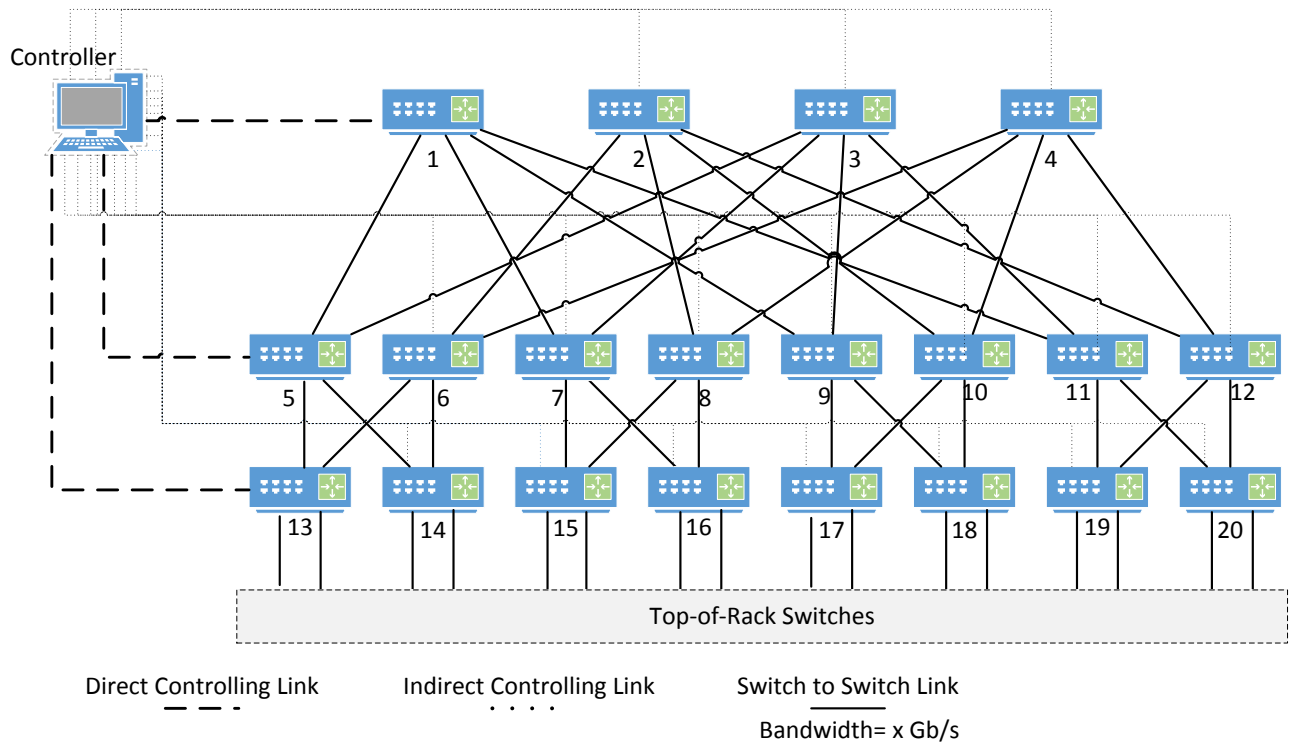

FIG. 12 SAMPLE SDN ENABLED NETWORK.

Fig. 13 shows the current state of the network in which the packet loss for these two flows is high (if the flows are based on UDP then the packet loss is 0.7x Gb/s). At this point, the controller decides to reroute the flows in a way that the network congestion in the link between S13 and S5 reduces. In the best case, the red flow will be rerouted similar to Fig. 14. Based on the mentioned reasons, the impact of SFTC on routing is a function of 1) number of rerouted flows and 2) network topology. As mentioned earlier, number of rerouted flows is illustrated in Fig. 9 and it is shown that the QNR is superior in comparison with conventional approaches like Shortest-Path.

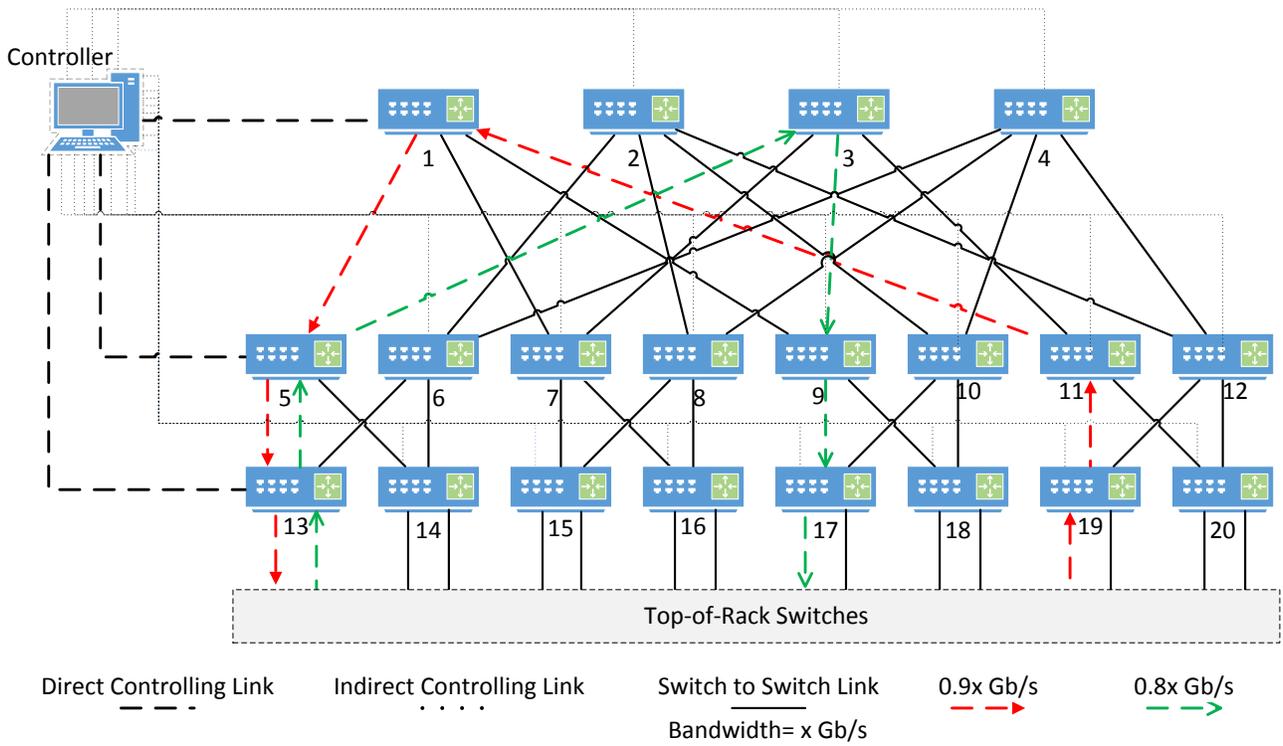

FIG. 13 CURRENT STATE OF THE NETWORK.

In Fig. 14, the controlling message is delivered to S4 and S19 with $3\tau$ second delay. As a result, the forwarding table of S19 is updated in $3\tau$ seconds after the controller sends the controlling messages. This means that the link $S5 \rightarrow S13$ remains in congestion and some parts of the packets that are sent from S19 in this time period ($3\tau$ second) will be lost. Consequently, the packet loss of red and green flows is $x - (F1 + F2)\ Gb/s$ which is in our case $0.7x * \tau\ Gb/s$.

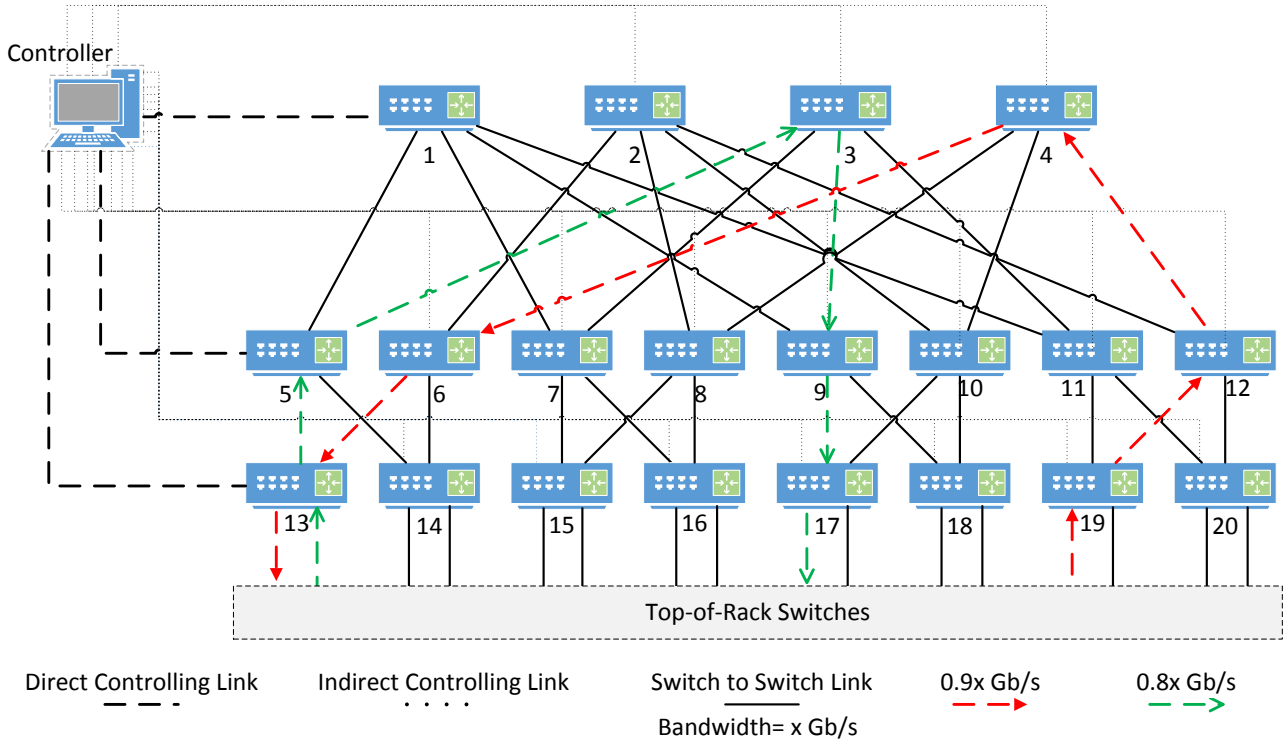

- The Impact of TCP Retransmission Timer on the Packet Loss

In order to evaluate the impact of SFTC on TCP retransmission timer, we consider the cases in which the delay of new route for a specified flow is sufficiently greater than the delay of the old route. As an example, consider that the current state of the network is as Fig. 15.

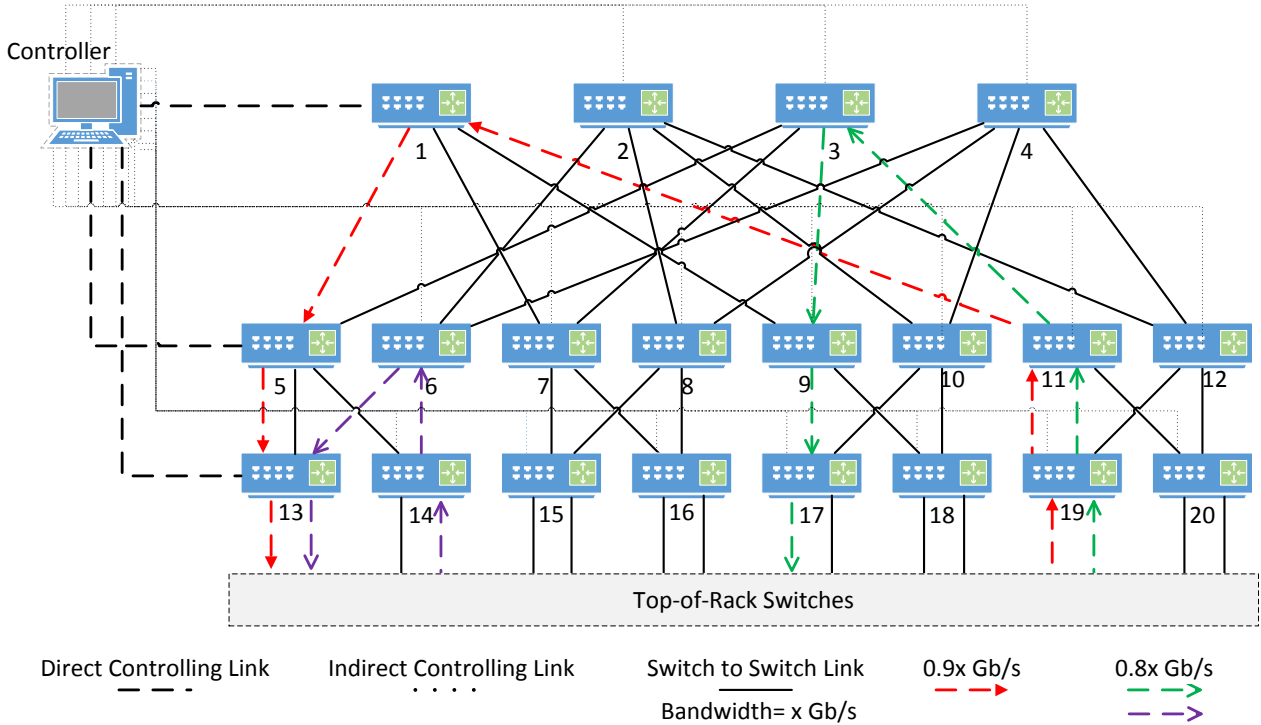

FIG. 15. TCP RETRANSMISSION EFFECT- CURRENT STATE OF THE NETWORK.

The state of the network after rerouting of the red flow is depicted in Fig. 16. As can be seen, the packets of red flow will force a delay of $2\tau$. Therefore, the TCP retransmission timeout is occurred and some packets will be lost. In real network, this effect is multiplied since the reconfiguration algorithm is used periodically. Briefly, if the number of rerouted flows is increased, the probability of packet loss increases.

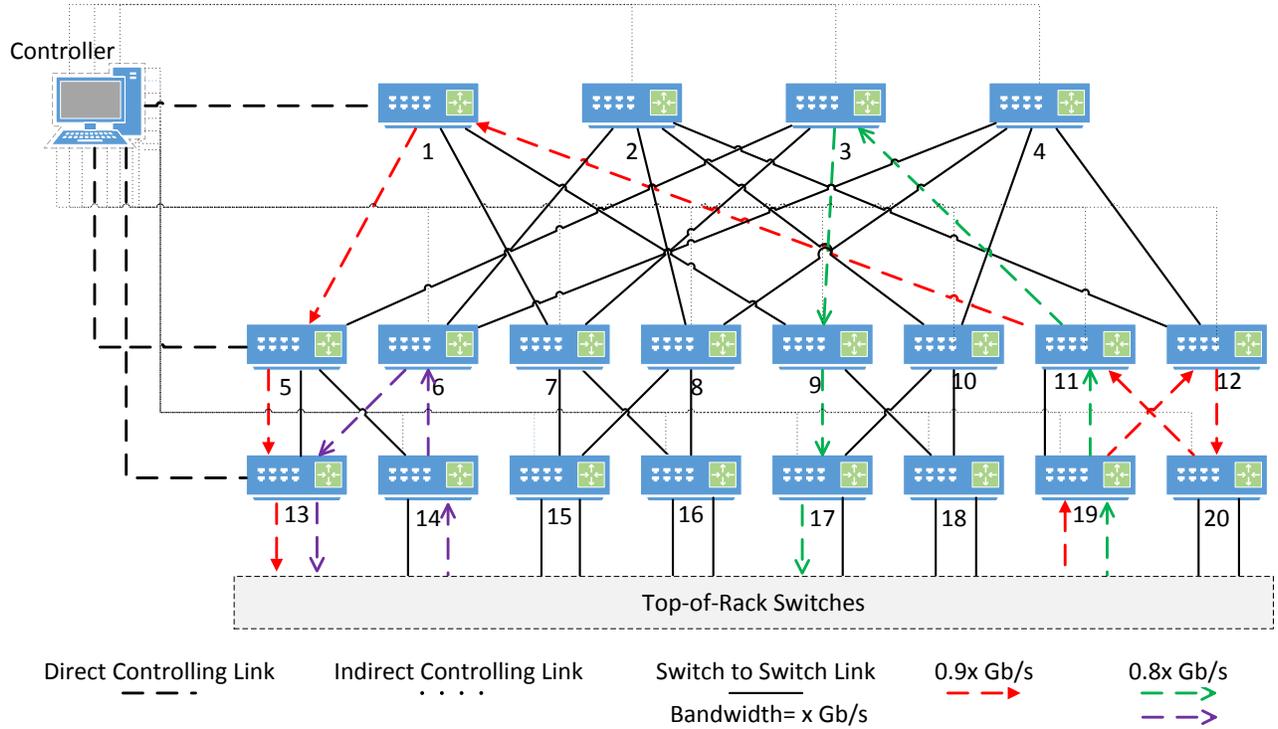

*FIG. 16. TCP RETRANSMISSION EFFECT- NEXT STATE OF THE NETWORK.*

### 6.3 Complexity Analysis

#### 6.3.1 QNR

In our emulation, in order to speed up the process of implementing QNR, we use CVX (a MATLAB toolbox) to solve our proposed algorithm. There are lots of other methods that are faster than CVX in solving binary linear programming problems such as the work proposed in [22] which has a computational complexity of $O(p^2 * n^4) \approx O(p^2 * k^6)$ where $p$ is the number of flows and $k$ is the order of fat-tree topology (in our emulation it is 4 and 6). Therefore, we can say that the computational complexity of solving the proposed scheme using [22] is $O(p^2 * k^6)$ in average. Additionally, we present the computational complexity of the worst case (brute force) in solving the problem.

QNR is a BLP problem, therefore, each link is labeled with $a_{if} \in \{0,1\}$ in which the variable $a_{if}$ determines whether the flow $f$ uses the link $i$ or not. Consequently, the complexity of brute force in finding the solution is depended on $(number\ of\ links) * (number\ of\ flows)$. In this way, the number of links is calculated in the fat-tree topology. In core layer, there are $\frac{k}{2} * \frac{k}{2}$ switches, each one has $k$ links, accordingly, there are $\frac{k}{2} * \frac{k}{2} * k$ links in the core layer. On the other hand, there are $k * \frac{k}{2}$ switches in the aggregation layer that uses two groups of $\frac{k}{2}$ links to connect to both core layer and access layer. As a result, there are $k * \frac{k}{2} * k$ variables that must be calculated. Similar to aggregation layer, there are $\frac{k}{2} * k$ switches in access layer. Since $\frac{k}{2}$ links of access switches are connected to client devices, we ignore these links that do not play role in routing, therefore, there are $\frac{k}{2} * k * \frac{k}{2}$ links in access layer. The total number of variables that must be calculated is: $k * \frac{k}{2} * \frac{k}{2} + k * k * \frac{k}{2} + \frac{k}{2} * \frac{k}{2} * k = k^3$

As the problem is in form of binary, the total states are $2^{k^3}$. On the other hand, for each state, all constraints must be satisfied which means that the computational complexity of checking the constraint violation must be considered. In this regard, to check the satisfaction of (8), one needs $n * p$ operation of adding $n$ parameters, therefore, the computational complexity is $n * p * n$.

Similarly, (2) and (7) need $n*p*n$ operations to be checked. Each of (3), (4), (5), and (6) need $n*p$ operations. Consequently, the total cost of validation of each state is: $3*n*p*n + 4*n*p = 3n^2p + 4np$.

Based on the fact that $n = k*k + k*\frac{k}{4} = \frac{5k^2}{4}$ we can say the cost of validation is equal to: $3\left(\frac{5k^2}{4}\right)^2 p + 4\left(\frac{5k^2}{4}\right)p = p\left(\frac{75k^4}{4} + 5k^2\right)$.

Consequently, the total computational complexity of brute force algorithm in finding the solution of QNR is $2^{k^3} * p\left(\frac{75k^2}{4} + 5k^2\right)$. As it can be seen, QNR is proper for medium and small size networks. In the next section, we show that the computational complexity of RQNR is dramatically lower than QNR which makes it more efficient for real world networks.

Since the version of shortest path which is used in this paper is QoS-aware, it is solved with the CVX toolbox. Therefore, although the computational complexity of Dijkstra is $O\left(\frac{5k^2}{4} + k^3\right)$, the computational complexity of QoS-aware Shortest Path cannot be evaluated precisely. Similar to QNR, the complexity of QoS-aware Shortest Path can be evaluated for brute force algorithm and it is equal to the computational complexity of QNR. It is notable that if we use [22] to solve the Shortest Path, the computational complexity is $O(p^2 * n^4) \approx O(p^2 * k^6)$ which is similar to the computational complexity of QNR.

### 6.3.2 RQNR

Since RQNR reduces the number of active flows, it cannot be compared with QNR from the objective function value perspective. Accordingly, the number of active flows versus the total number of flows in different cases is measured and illustrated in Fig. 17 and Fig. 18 to compare QNR and RQNR. The topology which is used in Fig. 17 and Fig. 18 is Fat-tree K=6 except Fig. 17A and Fig. 18A which are based on the fat-tree K=4 topology. The difference in these figures are about the traffic pattern which is explained by the parameter PL. A path satisfying all constraints is called a feasible solution. Based on our experiments, RQNR finds a feasible solution for all of our test cases. As it can be seen, by using the forwarding table compression technique, the number of active flows is reduced more than 90 percent in some cases. Increasing the number of flows, makes the forwarding table compression technique more efficient, while increasing the number of switches makes it less effective. Hence, the compression rate of the proposed algorithm in Fig. 17B and Fig. 17C (in compare with Fig. 17A) is decreased for the reason that the number of switches is increased. In Fig. 18 the forwarding table compression rate versus the total number of flows is depicted. As it can be seen, the rate of compression increases by increasing the number of flows, however, increasing the number of links reduces the compression rate.

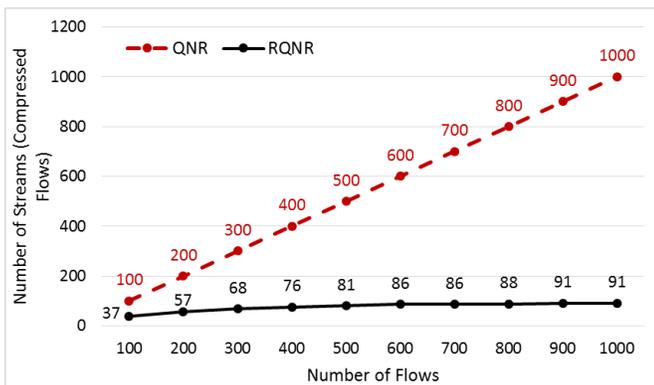

Fig. 17A) Fat-Tree K=4 PL=0.25

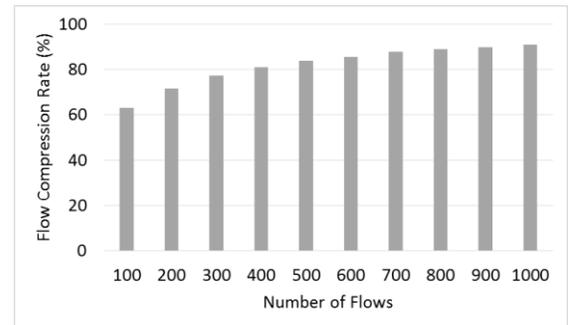

Fig. 18A) Fat-Tree K=4 PL=0.25

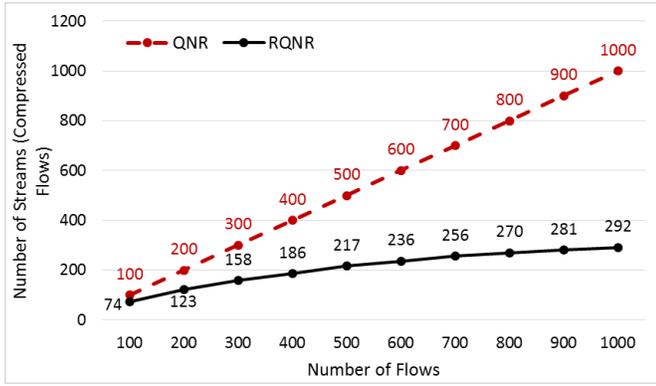

Fig. 17B) Fat-Tree K=6 PL=0.5

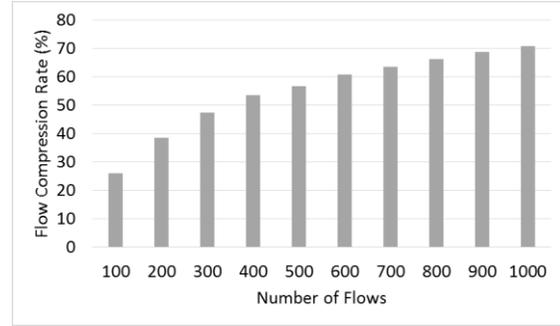

Fig. 18B) Fat-Tree K=6 PL=0.5

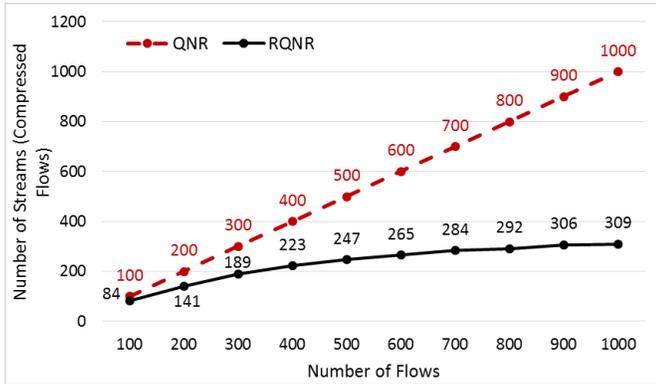

Fig. 17C) Fat-Tree K=6 PL=0.75

FIG. 17. COMPARISON OF QNR & RQNR (COMPRESSED FLOWS).

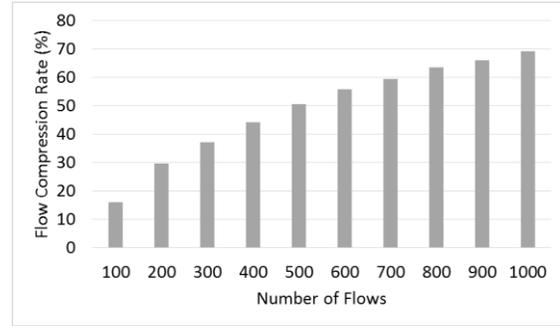

Fig. 18C) Fat-Tree K=6 PL=0.75

FIG. 18 COMPARISON OF QNR & RQNR (COMPRESSION RATE).

- Sub-optimality of RQNR

Although RQNR founds a feasible solution for each flow in all of our test cases, it is semantically sub-optimal and in some rare cases it cannot find a feasible solution. To show this, we illustrate an example in which RQNR could not find a feasible solution while QNR finds one. Consider the topology illustrated in Fig. 12 with four flows in the network. All links capacity are equal to 100 Mb/s and the flows are $(F1: S1 \rightarrow S5), (F2: S1 \rightarrow S3), (F3: S1 \rightarrow S4), (F4: S1 \rightarrow S5)$ where the corresponding rates are 95, 96, 4, 4 Mb/s, respectively. QNR schedules the network similar to Fig. 19 in which the flows QoS requirements are satisfied correctly.

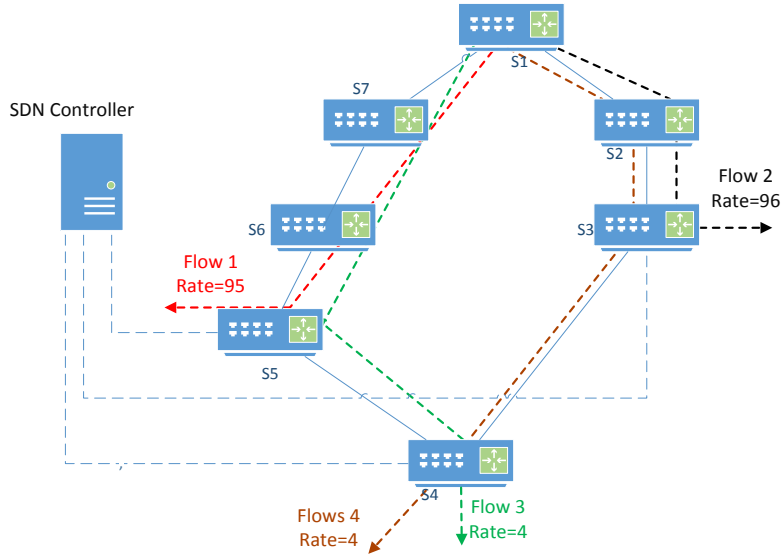

FIG. 19 ROUTING WITH QNR.

Since RQNR uses forwarding table compression, the output streams are $F1:(S1 \rightarrow S5), (F2:S1 \rightarrow S3), (FN:S1 \rightarrow S4)$ where the new rates are 96, 95, and 8 Mb/s, respectively. As it can be seen from Fig. 20, although there is a feasible solution, RQNR could not find a valid reallocation, i.e. the output violates a constraint and results in congestion in the network.

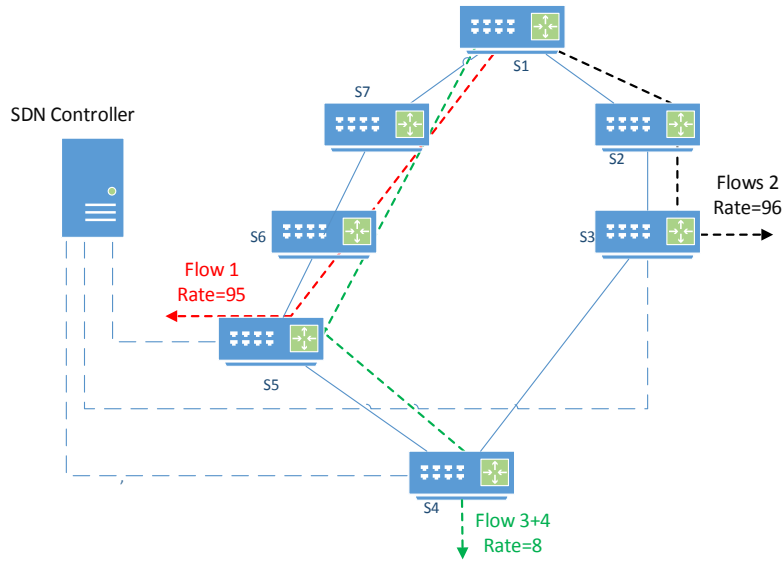

FIG. 20 ROUTING WITH RQNR.

### 6.3.3 Constraints

In order to show the impact of each constraint on the computational complexity of the proposed scheme, the algorithm is executed in the absence of each constraint. There are 22 switches and the topology is illustrated in Fig. 4. The number of flows is set to 100, 200, 300, and 400 and the execution time of the program without each constraint is averaged. Fig. 21 illustrates the percentage of the CPU time which is dedicated to each constraint. As it can be seen, the most time consuming part is related to the process of forcing flows to leave the source switch and enter to the destination one.

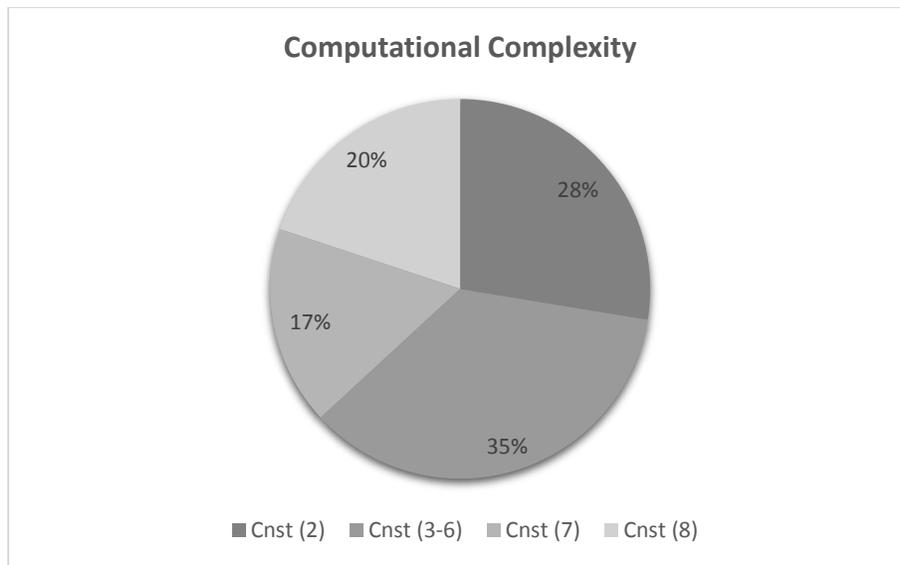

FIG. 21 COMPUTATIONAL COMPLEXITY OF EACH CONSTRAINT ON TOTAL COMPLEXITY.

This information about the computational complexity of constraints can be used as a guidance to relax the constraints. In this way, post processing (or preprocessing) is required, e.g., one can remove the constraints 3-6 and substitute a preprocessing instead of these computationally complex constraints.

7 CONCLUSION

In this paper, a novel QoS-aware resource reallocation algorithm, called QNR, was introduced. This algorithm reschedules the network in a way that imposes the minimal overhead in forwarding table updates. Due to the fact that QNR uses the SDN abilities for resource reallocation, it is proper for networks with dynamic traffics in which the resource partitioning leads to a high packet loss for big flows. The problem was mathematically formulated in the form of binary linear programming. In order to solve the corresponding optimization problem, a scheme was proposed. The performance of the proposed scheme was compared with the shortest-path via three measures 1) network reconfiguration overhead (SFTC), 2) delay, and 3) packet loss. In all of our test cases, the SFTC and delay measures improved more than 90 and 60 times, respectively. Additionally, in order to completely evaluate the effect of QNR on network throughput, the packet loss was compared in a simple topology which showed a dramatic decrement in the packet loss. Finally, QNR and RQNR were compared from time complexity perspective. In this way, it is shown that the proposed forwarding table compression technique (which is used in RQNR) decreases the overhead of FTs update by decreasing the number of active flows (the forwarding table compression rates were between 15% and 90%). Future works will be dedicated on proposing heuristic methods for hybrid networks (combination of conventional and OpenFlow enabled switches).